\documentclass[prd,aps,twocolumn,nofootinbib,superscriptaddress,eqsecnum,floatfix,preprintnumbers,amsmath,amssymb,longbibliography]{revtex4-2}
\usepackage{csquotes}

\makeatletter
\renewcommand{\p@subsection}{}
\renewcommand{\p@subsubsection}{}
\makeatother

\usepackage{graphicx}
\usepackage{epsfig}
\usepackage[dvipsnames]{xcolor}
\usepackage[utf8]{inputenc}
\usepackage{stmaryrd}
\usepackage{mathrsfs}
\usepackage{mathalfa}
\usepackage{accents}
\usepackage{enumitem}
\usepackage[normalem]{ulem}

\usepackage[colorlinks=true,
            citecolor=green,
            linkcolor=red,
            filecolor=cyan,
            urlcolor=magenta,
            backref=false]{hyperref}

\usepackage{orcidlink}

\newcommand{\pa}{\partial}

\newcommand{\be}{\begin{equation}}
\newcommand{\ee}{\end{equation}}
\newcommand{\bea}{\begin{eqnarray}}
\newcommand{\eea}{\end{eqnarray}}
\newcommand{\ba}{\begin{equation}\begin{aligned}}
\newcommand{\ea}{\end{aligned}\end{equation}}
\newcommand{\beg}{\begin{gather*}}
\newcommand{\eng}{\end{gather*}}
\newcommand{\hh}{,\hspace{0.5cm}}

\newcommand{\n}[1]{\label{#1}}
\newcommand{\ins}[1]{{\mbox{\tiny #1}}}

\newcommand{\MC}[1]{{\mathcal{#1}}}
\newcommand{\CAL}{\mathcal}

\newcommand{\bs}{\begin{split}}
\newcommand{\es}{\end{split}}

\begin{document}

\title{Quasitopological Gravity with Matter: Modified Double-Copy Approach}

\author{Valeri P. Frolov \,
\orcidlink{0000-0002-8414-5965}%
}
\email[]{vfrolov@ualberta.ca}
\affiliation{Theoretical Physics Institute, Department of Physics,
University of Alberta,\\
Edmonton, Alberta, T6G 2E1, Canada
}


\begin{abstract}
We extend the recently proposed modified double-copy formalism to
quasitopological gravity (QTG) coupled to matter.
For spherically symmetric configurations, the QTG field equations in $D-$dimensional curved spacetime
with a broad class of matter sources are mapped to equations for an
auxiliary nonlinear gauge field in a flat $(D+1)$-dimensional
spacetime. The nonlinear electrodynamics governing this auxiliary
field is determined entirely by the generating function $h(p)$ that
specifies the QTG model, while the corresponding current is determined
by the matter stress-energy tensor. Restricting the auxiliary solution
to a $D$-dimensional hyperplane and applying the modified double-copy
prescription yields the Kerr--Schild metric solving the QTG equations.
We show that Maxwell and nonlinear electrodynamics, as well as a broad
class of spherically symmetric Yang--Mills fields, provide physical
matter sources compatible with this construction. 
In the absence of null-matter fluxes, the corresponding solutions possess an additional Killing vector.
This means that they satisfy a generalized Birkhoff
theorem, whereas null charged currents naturally generate
Vaidya-type solutions. In the Einstein limit, $h(p)=p$, the auxiliary
nonlinear electrodynamics reduces to Maxwell theory.

\hfill    Alberta Thy 4-26

\end{abstract}

\maketitle

\section{Introduction}

A broad class of important solutions of the Einstein equations can be written in the Kerr--Schild form \cite{KerrSchild1965}
\begin{equation}
\label{KS}
ds^2 = ds_0^2 + \Phi \,(k_{\mu}dx^{\mu})^2 \, ,
\end{equation}
where $ds_0^2$ is the flat spacetime metric and $k_{\mu}$ is a shear-free null congruence. The vector field $k^{\mu}$ is null with respect to both the background metric $ds_0^2$ and the full metric $ds^2$.
A remarkable property of the Kerr--Schild ansatz is that certain exact solutions of the highly nonlinear Einstein equations can be generated from a class of solutions of the linear Maxwell equations through the classical double-copy correspondence, provided the Kerr--Schild vector is geodesic. In this construction, the null vector $k^{\mu}$ is a principal null eigenvector of the electromagnetic field strength tensor $F_{\mu\nu}$, while the scalar function $\Phi$ is determined by the corresponding electromagnetic potential
(see e.g. \cite{Bern_2010,Monteiro_2014,Luna_2015,Bah:2019sda}. See also review articles \cite{White_2018,bern2019duality,bern2022sagex} and references therein).

It is well known that both the Schwarzschild and Kerr solutions, although belonging to the Kerr--Schild class \cite{Visser:2007fj}, possess curvature singularities in their interiors. This feature is widely regarded as an indication of the incompleteness of classical general relativity. Consequently, numerous attempts have been made to modify Einstein's theory in order to resolve these singularities. Among the most promising recent developments is the so-called \emph{quasitopological gravity} (QTG) model, which provides a novel higher-curvature extension of general relativity \cite{Bueno:2020gq,Bueno:2019ltp,Bueno:2022res,Oliva_2010,Hennigar:2017ego,Myers:2010ru,Moreno:2023rfl,QT_BH,rbh_pfkz,PinedoSoto:2025hel, Bueno:2025tli, Bueno:2025zaj, Frolov:2025ddw, Bueno:2026dln,Sueto:2026epz,Borissova:2026krh,Borissova:2026wmn,Borissova:2026rbi,Bueno:2025qjk}
and admits regular black-hole solutions
\cite{QT_BH,bueno2025nonpolynomial, Myers:2010ru, rbh_pfkz, Bueno:2025tli, Bueno:2025zaj,PinedoSoto:2025hel,Bueno:2026dln}.

In this paper we show that the Kerr--Schild formalism naturally extends to QTG and can be used to construct new classes of exact regular black-hole solutions.

The quasitopological gravity (QTG) theory is formulated by extending the Einstein--Hilbert action with an infinite sequence of higher-curvature contributions,
\begin{equation}
\label{QTA}
\begin{split}
S_{\rm QTG} &= \frac{1}{2\varkappa}\int d^D x\,\sqrt{-g}\,L_{\rm QTG}\,,\\
L_{\rm QTG} &= R + \sum_{j=2}^{\infty} \alpha_j \ell^{2(j-1)} Z_j \, .
\end{split}
\end{equation}
Here $\varkappa=8\pi G^{D}$, where $G^{D}$ denotes the gravitational coupling constant in $D$ spacetime dimensions.
The quantities $Z_j$ are curvature invariants constructed as polynomials of degree $j$ in the Riemann tensor and its contractions. They are chosen so that, upon imposing spherical symmetry, the resulting gravitational field equations remain second order in derivatives despite the presence of arbitrarily high powers of the curvature.

The explicit expressions for the invariants $Z_j$, together with the recursive procedure used to generate them, are given in \cite{Bueno:2020gq}. The parameter $\ell$, which has the dimension of length, determines the scale at which higher-curvature corrections become important, while the dimensionless coefficients $\alpha_j$ specify a particular member of the QTG family of theories.

A distinctive feature of this construction is that the series in Eq.~\eqref{QTA} need not be truncated. For suitable choices of the coefficients $\alpha_j$, the complete infinite series admits spherically symmetric solutions describing regular black holes, for which the curvature remains finite throughout the spacetime. It is precisely this property that makes QTG particularly attractive. It opens the possibility of studying the formation and evolution of black holes without encountering curvature singularities in their interiors. It also provides a framework for revisiting long-standing fundamental problems of black hole physics in Einstein gravity, such as mass inflation and information loss.

An important feature of QTG is that the sum in Eq.~\eqref{QTA} is generally taken over an infinite tower of curvature invariants. The couplings $\alpha_j$ are conveniently encoded in a generating function \cite{Bueno:2025gjg}
\be
h(p)=p+\sum_{j=2}^{\infty} c_j \ell^{2(j-1)}p^j\hh c_j=\dfrac{D-2j}{D-2}\alpha_j\, .
\ee
The variable $p$ appearing in this relation is the so-called \emph{primary curvature invariant}. Its explicit definition will be given below. The function $h(p)$ cannot be chosen arbitrarily. Rather, it must be analytic and invertible over the relevant domain of $p$. Under these conditions, the spherically symmetric field equations possess a unique physical branch of solutions. Moreover, appropriate choices of $h(p)$ give rise to asymptotically flat regular black-hole spacetimes in which all curvature invariants remain finite everywhere \cite{Bueno:2020gq}.

A principal objective of this paper is to demonstrate that the modified double-copy formalism recently proposed in \cite{Frolov:2025ddw} provides a powerful framework for constructing solutions of quasitopological gravity coupled to matter. The key distinguishing feature of this approach is that it employs a $(D+1)$-dimensional flat spacetime $M^{D+1}$ to generate solutions of the QTG field equations in a curved $D$-dimensional spacetime $\CAL{M}^D$.
The construction is based on solving the equations of motion for an auxiliary gauge field in $M^{D+1}$, governed by a nonlinear electrodynamics in the presence of an appropriately chosen electric current. In the Einstein limit, corresponding to $h(p)=p$, these equations reduce to the linear Maxwell equations. In contrast, for a generic QTG model, the auxiliary gauge field is governed by a nonlinear electrodynamics whose Lagrangian is uniquely determined by the generating function $h(p)$.

The paper is organized as follows. In Section\,\ref{Sec2}, we discuss equations of QTG in the presence of matter field and formulate a special ansatz for the stress-energy tensor of the spherically symmetric matter distribution   considered in this paper.
In Section\, \ref{Sec3} it is demonstrated that a modified double-copy formalism allows one to obtain the solutions for the corresponding gravitational field in QTG. In Section\, \ref{Sec4}  it is demonstrated that the described modified double-copy formalism allows one to obtain solutions for QTG coupled with non-linear electrodynamics and Yang-Mills fields. Finally, in Section\,\ref{Sec5}, we summarize our results and discuss possible directions for future research.
 In the paper we use units in which $c=1$ and sign convention adopted in the book \cite{MTW}.

\section{QTG dilaton 2D action and field equations}\label{Sec2}

\subsection{Spherically reduced QTG equations}

We first summarize the geometrical conventions employed in the spherical reduction of the QTG theory. Let $\CAL{M}^D$ denote a $D$-dimensional curved spacetime endowed with the metric $g_{AB}$,
\be \n{DDD}
ds^2=g_{AB} dX^{A} dX^B\hh A,B =0,1,\ldots, D-1 \, .
\ee
We restrict attention to geometries that split into a two-dimensional orbit space and a round $(D-2)$-sphere. The corresponding warped-product ansatz is
\be\n{GO}
ds^2=\gamma_{\mu\nu}(x) dx^{\mu} dx^{\nu}+r^2(x)
d\omega^2_\ins{D-2}\, ,
\ee
Here
\be \n{OO}
d\omega^2_\ins{D-2}=\omega_{ij} dy^i dy^j\,
\ee
denotes the standard line element on the unit sphere $S^{D-2}$. Greek indices  $(\mu,\nu,\ldots)$ refer to the two-dimensional orbit space and take the values $0,1$, whereas $(i,j,\ldots)$ label the angular directions $2,\ldots,D-1$. The area of the unit $(D-2)$-sphere will be written as $\Omega_\ins{D-2}$, with
\be\n{OM}
\Omega_\ins{D-2}=\frac{2\pi^{(D-1)/2}}{\Gamma\big(\frac{D-1}{2}\big)} .
\ee

Any symmetric rank-two tensor compatible with spherical symmetry admits a decomposition into orbit-space and angular pieces. With the normalization adopted here, this decomposition is written as
\be\n{PPPP}
P_{A}{}^{B}=\dfrac{1}{r^{D-2}}\Big[
\delta^{\mu}_{A}\delta^{B}_{\nu}\,\mathcal{P}_{\mu}{}^{\nu}
+
\delta^{i}_{A}\delta^{B}_{i}\frac{\mathcal{P}}{D-2}
\Big]
\ee
where $\mathcal{P}_{\mu\nu}$ is a symmetric tensor intrinsic to the two-dimensional orbit space and $\mathcal{P}$ is a scalar. The factor $1/(D-2)$ ensures that $\mathcal{P}$ is precisely the rescaled trace of the angular sector, $\mathcal{P}=r^{D-2}{P}^{i}{}_{i}$.

The covariant derivative associated with the full $D$-dimensional metric is denoted by $\nabla_A$. Covariant differentiation with respect to $\gamma_{\mu\nu}$ is indicated by a semicolon. If the tensor $P_{AB}$ obeys the conservation law
\be
\nabla_{B}P^{B}{}_{A}=0,
\ee
then its reduced components satisfy
\be \n{CONS2}
\CAL{P}^{\nu}_{\ \mu ;\nu}
=
\frac{1}{r}\,
r_{;\mu}\,\mathcal{P}.
\ee

A general spherically symmetric geometry possesses four algebraically independent scalar curvature invariants \cite{NARLIKAR}. The full curvature tensor is determined by the primary curvature invariant $p$
\be
p=\dfrac{1-(\nabla r)^2}{r^2}\hh (\nabla r)^2=\gamma^{\mu\nu}r_{,\mu}r_{,\nu}\, ,
\ee
 the two-dimensional curvature $\CAL{R}$, and the tensor $r_{;\mu\nu}/r$, together with the metric.

After taking integral in \eqref{QTA}
over the angular variables, the QTG action reduces to
\be\n{AQTG}
S_\ins{QTG}=B\CAL{S}_\ins{QTG}[\gamma,r] ,
\ee
where
\ba\n{AQTG1}
\CAL{S}_\ins{QTG}&=\int d^2 x \sqrt{|\gamma|}\mathcal{L}, \\
\mathcal{L}&=\frac{1}{D-2}r^{D-2} L_\ins{QTG}
\, .
\ea
The constant $B$ collects the contribution from the angular integration and is given by
\be
B=\frac{(D-2)\Omega_\ins{D-2}}{16\pi G_D}=\frac{(D-2)\Omega_\ins{D-2}}{2\kappa}\, ,
\ee
Here $G_D$ denotes the gravitational coupling in $D$ dimensions. The second equality introduces the parameter $\kappa$; in four dimensions it reduces to $\kappa=8\pi G$.

For the QTG theory under consideration, the reduced Lagrangian density $\CAL{L}$ can be expressed as follows (see Eq.~(11) of \cite{Bueno:2025gjg}):
\ba\n{LQTG}
\MC{L}=&G_2(r,f)-\Box r G_3(r,f)+G_4(r,f) \CAL{R}
\\
&-2\big(\partial_f G_4(r,f)\big)\big[(\Box r)^2-r^{;\alpha\beta}r_{;\alpha\beta}\big]\, ,
\ea
with
\be
\begin{split}
&G_2 =r^{(D-2)}(D-1)h -2r^{(D-2)}{p} h' \, ,  \\
&G_3=  2r^{(D-3)}h'\, ,  \\
&G_4=  -\frac{1}{D-2}r^{(D-2)}\lambda\, ,\\
&\lambda({p})=\frac{D-2}{2}{p}^{(D-2)/2}\int d {p}\, {p}^{-D/2} h'({p})\, .
\end{split}
\ee
Here and later $\Box$ is a two-dimensional "Box" operator and 
\be
f=(\nabla r)^2=\gamma^{\mu\nu}r_{,\mu}r_{,\nu}\, .
\ee
The function $h=h(p)$ determines the particular QTG model through its dependence on the primary curvature invariant $p$. We keep $h(p)$ unspecified at this stage and introduce a concrete choice only when it becomes necessary. A prime always denotes differentiation with respect to $p$.

The spherically reduced QTG action \eqref{AQTG1} can in fact be regarded as a particular form of two-dimensional dilaton gravity. In the presence of matter, the corresponding reduced matter action must be added to the gravitational action. The field equations are obtained by varying the total reduced action with respect to the two-dimensional metric $\gamma_{\mu\nu}$ and the dilaton field $r$. These equations inherit the same spherical decomposition as in \eqref{PPPP}.
To cast the equations in a form analogous to the standard Einstein equations, we place the gravitational contributions on the left-hand side and the matter stress-energy tensor on the right-hand side. We denote the orbit-space and angular components of the gravitational sector by $\CAL{G}_{\mu\nu}$ and $\CAL{G}$, respectively, and the corresponding components of the matter stress-energy tensor by $\CAL{T}_{\mu\nu}$ and $\CAL{T}$. The reduced field equations for QTG coupled to matter then take the form
\be \n{GEQ}
\CAL{G}^{\mu\nu}=\frac{2\kappa}{D-2}\CAL{T}^{\mu\nu}\hh \CAL{G}=\frac{2\kappa}{D-2}\CAL{T}\, .
\ee
Applying the reduced conservation identity \eqref{CONS2} separately to the gravitational and matter sectors gives
\be\label{Bianchi0}
\mathcal{G}_{\mu\alpha}{}^{;\alpha}=\frac{1}{r}r_{;\mu} \CAL{G}\hh
\mathcal{T}_{\mu\alpha}{}^{;\alpha}=\frac{1}{r}r_{;\mu} \CAL{T}\, .
\ee
Suppose that the first set of the gravitational field equations in \eqref{GEQ} is satisfied. Then the Bianchi identities \eqref{Bianchi0} ensure that the second (angular) equation in \eqref{GEQ} is automatically satisfied as well.

\subsection{QTG equations in $(v,r)-$coordinates}

Let us consider a $2D$ part of the metric \eqref{GO} in more details. Let $e_{\mu\nu}$ be a $2D$ unit antisymmetric tensor. Denote
\be
\xi^{\mu}=e^{\mu\nu}r_{,\nu}\, .
\ee
There is a sign ambiguity in the definition of $e_{\mu\nu}$, which we fix by requiring the vector $\xi^\mu$ to be future-directed. It is then straightforward to verify that
\be
\xi^2=-f\, ,
\ee
and the vector
\be
k_{\pm}^{\mu}=\xi^{\mu}\pm r^{,\mu}
\ee
is null. One also has
\be
k_{\pm}^{\mu}r_{,\mu}=\pm f\, .
\ee
This means  that for sign plus in the domain where $f>0$ the radius increases along the null ray with a future directed tangent vector $k^{\mu}$. We call such rays out-going. For opposite choice of the sign, the rays are called in-coming. If in a some domain the gradient of radius does not vanish, the out-going null rays do not intersect and form a foliation of the null rays in this domain. Similarly, one has another foliation formed by in-coming rays\footnote{
In fact these rays satisfy the equation
\[
k_{\pm}^{\nu}\nabla_{\nu}k_{\pm}^{\mu}=\pm \Box r k_{\pm}^{\mu}\, ,
\]
and hence they are geodesics.
}.
In what follows we focus on the in-coming null rays. We denote by $v$ a scalar function which takes constant value of each of the in-coming ray and can be used to 'enumerate' the rays. Evidently, this coordinate is defined up to a possible reparametrization $v\to A(v)$. Using the function $r=r(x)$ as a second coordinate, one gets a coordinate chart $(v,r)$ in which the metric \eqref{GO} takes the form
\be \n{NFR}
ds^2=-N^2 f dv^2+2 N dv dr +r^2 d\omega^2_\ins{D-2}\, .
\ee
We now consider the metric in the above form, allowing the functions $f$ and $N$ to depend on both $v$ and $r$, i.e., $f=f(v,r)$ and $N=N(v,r)$.

For this metric, after lengthy but straightforward calculations, one obtains the tensor $\CAL{G}_{\mu\nu}$ and finds that (see \cite{Frolov:2026tft} for more details)
\be \n{EQQ}
\begin{split}
&\mathcal{G}_{vv}+N f \mathcal{G}_{vr}=N H_{,v}\, ,\\
&\mathcal{G}_{vr}=-N H_{,r}\, ,\\
&\mathcal{G}_{rr}=2r^{D-3}\frac{N_{,r}}{N} h'({p})\, .
\end{split}
\ee
Here, $h=h(p)$ is a function of the primary curvature invariant. It specifies the particular QTG model under consideration. For the time being, we leave this function arbitrary. We also introduce the notation
\be
H=r^{D-1}h\, .
\ee

\subsection{Stress-energy tensor ansatz}

Let us assume that the gravitational field is generated by matter distribution respecting the imposed symmetry of the spacetime.
Its stress-energy is of the form
\be\n{TTTT}
T_{A}{}^{B}=\dfrac{1}{r^{D-2}}\Big[
\delta^{\mu}_{A}\delta^{B}_{\nu}\,\mathcal{T}_{\mu}{}^{\nu}
+
\delta^{i}_{A}\delta^{B}_{i}\frac{\mathcal{T}}{D-2}
\Big]
\ee
Following the approach of \cite{Frolov:2026tft}, we make the additional assumption that $\mathcal{T}_{\mu}{}^{\nu}$  can be written in the form
\be \n{SETT}
\mathcal{T}_{\mu\nu}=\tau \gamma_{\mu\nu}+\sigma k_{\mu} k_{\nu}\, .
\ee
We now choose a new normalization of the ingoing null direction
\be
k_{\mu}=v_{,\mu}\, .
\ee
The term proportional to $\sigma$ describes the contribution of null-matter fluxes. In the absence of such fluxes, solutions of the QTG field equations obey a generalized Birkhoff theorem and there exists an additional Killing vector.  In a domain where it is timelike a solution is static (see, e.g., \cite{Frolov:2026tft}).

Let us first note that for the metric \eqref{NFR} we have $\gamma_{rr}=0$, and therefore $\CAL{T}_{rr}=0$.The $(r,r)$ gravitational field equation is $\CAL{G}_{rr}=0$. We assume that $dh/dp\ne 0$. Then this equation implies that $N_{,r}=0$, and the metric function $N$ depends only on the coordinate $v$. Since this dependence can be eliminated by a trivial redefinition (rescaling) of the coordinate $v$, one may always choose the gauge $N=1$. From now on, we shall work in this gauge.
In this gauge the metric \eqref{NFR} can be presented in the form
\be\n{ss0}
ds^2=ds_0^2+r^2 p (k_{\mu}dx^{\mu})^2\, ,
\ee
where
\be \n{NFR0}
ds_0^2=- dv^2+2 dv dr +r^2 d\omega^2_\ins{D-2}\, ,
\ee
is the flat metric in $D-$dimensional Minkowski spacetime $M^{D}$ written in spherically symmetric null coordinates. In standard Cartesian coordinates, $X^{A}=(T,X_1,\ldots , X_{D-1})$, it has the form
\be \n{CAR}
ds_0^2=-dT^2+(dX_1)^2+\ldots +(dX_{D-1})^2\, ,
\ee
where
\be
T=v-r\hh r^2=X_1^2+\ldots +X_{D-1}^2\, .
\ee

The stress-energy tensor \eqref{SETT} contains three arbitrary functions of $(v,r)$, $\tau(v,r)$, $\sigma(v,r)$ and $\CAL{T}(v,r)$.
The conservation equation \eqref{Bianchi0} implies
\be \n{EEQQ}
\sigma_{,r}=-\tau_{,v}\hh
\tau_{,r}=r^{-1}\CAL{T}\, .
\ee
The first of these equations establishes a relation between $\tau$ and $\sigma$, whereas the second determines $\mathcal{T}$ for a given function $\tau$. Let us emphasize that the metric function $f$ does not enter either of these equations. Consequently, the stress-energy tensor \eqref{TTTT}--\eqref{SETT} is conserved not only in the curved spacetime with metric $ds^2$, but also in the flat spacetime with metric $ds_0^2$.

Using \eqref{EQQ} and the expression for the adopted stress-energy tensor \eqref{SETT} the QTG field equations can be written in the form
\be \n{HHVR0}
H_{,v}=\frac{2\kappa}{D-2}\sigma \,\hh
H_{,r}=-\frac{2\kappa}{D-2}\tau\, ,
\ee
where $H=r^{D-1}h(p)$. The first of the conservation equations \eqref{EEQQ} provides the integrability condition for the above system of equations. As a result, these equations uniquely determine $H$  in a simply connected domain of $(v,r)-$plane up to an additive constant.

Consider a two-dimensional domain in the $(v,r)$ plane and choose an arbitrary reference point $(v_0,r_0)$ within it. If the value
\[
H_0=H(v_0,r_0)
\]
is known, then the function $H(v,r)$ can be obtained by integrating \eqref{HHVR0} along any curve connecting the points $(v_0,r_0)$ and $(v,r)$. Owing to the integrability condition, the result of this integration is independent of the particular choice of the integration path.

After determining the function $H=H(v,r)$, one obtains
\[
h(p)=\frac{H(v,r)}{r^{D-1}}.
\]
By inverting this relation, one finds $p=p(v,r)$. Substituting this result into \eqref{ss0} then yields the metric representing the desired solution of the QTG field equations in the presence of the matter source \eqref{SETT}.

\section{Modified double-copy method of solving QTG equations}\label{Sec3}

In this section we apply the modified double-copy formalism proposed in
\cite{Frolov:2025ddw} to quasitopological gravity (QTG) in the presence of a matter
source.
Within this framework, solutions of the QTG field equations on the
curved $D$-dimensional spacetime $\mathcal{M}^D$ are obtained
indirectly. The construction begins by solving the equations of a
nonlinear electrodynamics theory in the flat
$(D+1)$-dimensional spacetime $M^{D+1}$.\footnote{It should be
emphasized that the nonlinear electromagnetic and Yang--Mills fields
introduced later in section~\ref{Sec4} represent the physical matter sources
coupled to QTG. By contrast, the gauge field $\mathcal{A}_a$ considered
in the present section serves only as an auxiliary field used in the
modified double-copy construction. Whenever it is necessary to
distinguish it from the physical gauge fields, we refer to
$\mathcal{A}_a$ as the \emph{auxiliary gauge field}.}

\begin{figure}[!htb]%
    \centering
 \includegraphics[width=0.25\textwidth]{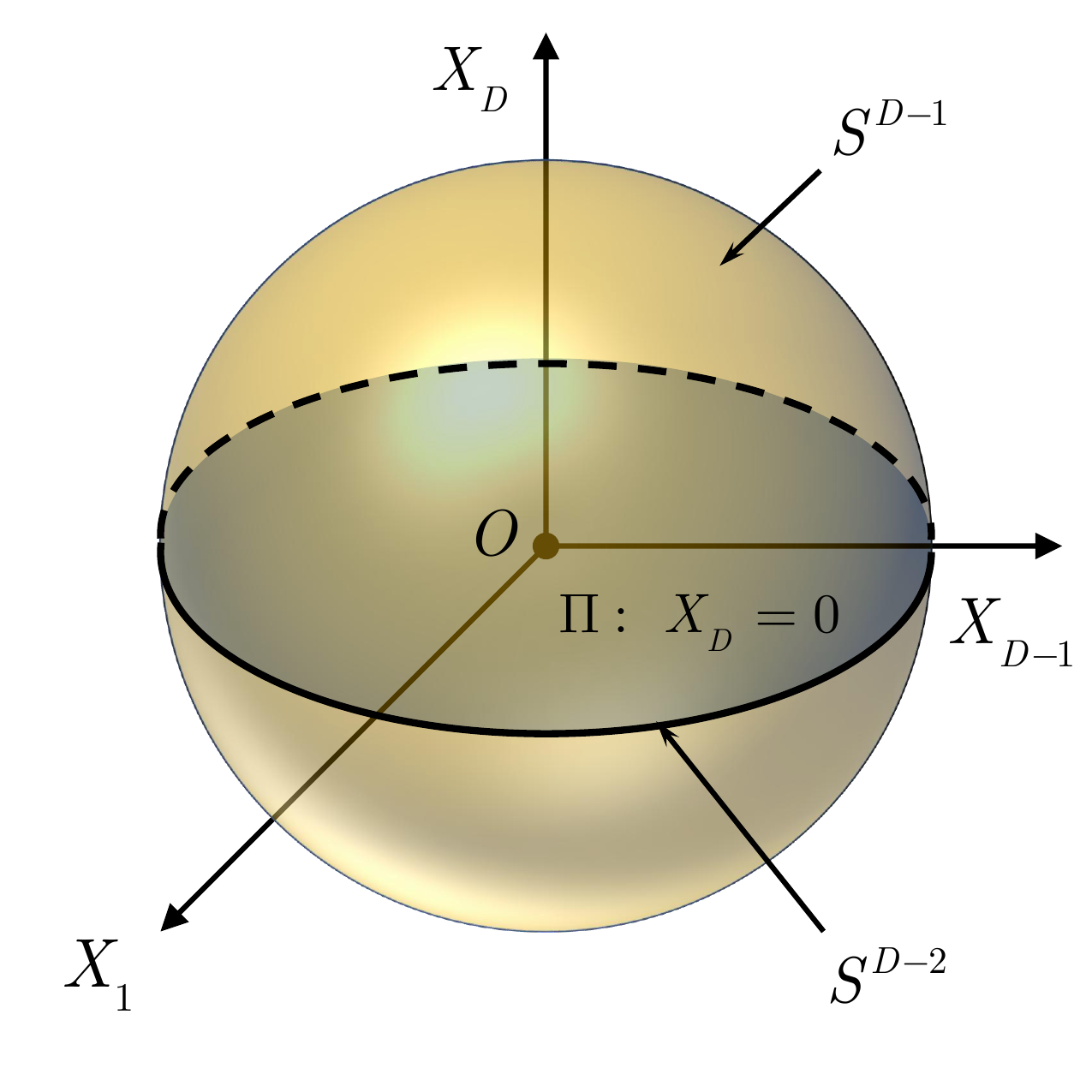}\\[0pt]
\caption{Schematic illustration of the modified double-copy construction.
A constant-time slice, $T=\mathrm{const}$, of the flat
$(D+1)$-dimensional spacetime $M^{D+1}$ is shown together with the
hyperplane $\Pi:\,X_D=0$, identified with the auxiliary spacetime
$M^D$. A sphere $S^{D-1}$ of radius $R$ intersects $\Pi$ in a sphere
$S^{D-2}$ of the same radius. The nonlinear electrodynamics equations
are first solved in $M^{D+1}$, the resulting gauge field is then
restricted to $\Pi$, and the modified double-copy prescription maps it
to a Kerr--Schild metric, yielding a solution of the quasitopological
gravity equations on the curved spacetime $\mathcal{M}^D$.}
    \label{F1}
\end{figure}

The resulting electromagnetic field is then restricted to a
$D$-dimensional hyperplane $M^D$. The induced gauge field is
subsequently used as the input for the modified double-copy
prescription, which reconstructs the corresponding Kerr--Schild metric
and thereby yields the associated solution of the QTG field equations on
the curved spacetime $\mathcal{M}^D$. The overall construction is
illustrated schematically in Fig.~\ref{F1}.

To this end we start with a $(D+1)$ Minkowski flat spacetime, which we denote by $M^{D+1}$. We denote by $X^{a}$ ($a=0,1,\ldots D$) Cartesian coordinates in it $X^a=(T,X_1,\ldots,X_{D-1},X_D)$ and write the metric in the form
\be
dS_0^2=-dT^2+(dX_1)^2+\ldots + (dX_{D-1})^2+(dX_D)^2\, .
\ee
We denote
\be
R^2=X_1^2+\ldots +X_{D-1}^2+X_D^2\hh
V=T+R\, .
\ee
We also introduce a null vector $K^{a}$ defined by $K_a=V_{,a}$.

We define the $D$-dimensional subspace specified by $X_D=0$ as the \emph{equatorial plane} and denote it by $\Pi$ (see Fig.~\ref{F1}). The metric induced on $\Pi$ coincides with the flat metric $ds_0^2$ given in \eqref{CAR}. It then follows immediately that
\be
R\big|_{\Pi}=r\hh V\big|_{\Pi}=v\hh K_{A}\big|_{\Pi}=k_{A} \, ,
\ee
where $\big|_{\Pi}$ denotes the pullback (restriction) of the corresponding object  to the equatorial plane $\Pi$.

For obtaining a solution of QTG with a matter source we consider a non-linear electrodynamics in a flat $(D+1)-$dimensional spacetime $M^{D+1}$ with the action of the form
\be\n{NLM}
\begin{split}
&W=\int d^{D+1}{X} L(\CAL{E})-\int d^{D+1} {X}  \CAL{J}^{a} \CAL{A}_{a}\, ,\\
&\CAL{E}^2=-\dfrac{1}{2} \CAL{F}_{ab}\CAL{F}^{ab}
\, .
\end{split}
\ee
The latter relation defines the electric field $\CAL{E}$ up to a sign which we specify later.
We assume that expansion of $L(\CAL{E})$ for small $\CAL{E}$ has the form
\be
L(\CAL{E}) =\dfrac{1}{2}\CAL{E}^2+\ldots\, ,
\ee
where dots denote higher in $\CAL{E}$ terms.

We shall search for spherically symmetric solutions of this theory. For this case both vectors $\CAL{J}^{a}$ and  $\CAL{A}_{a}$ depending on $(V,R)$ have two non-vanishing components, which in $(V,R)$ coordinates are $(\CAL{J}^{V},\CAL{J}^{R})$ and $(\CAL{A}_{V},\CAL{A}_{R})$, respectively. As a result, the field strength tensor $\CAL{F}_{ab}$ has only one non-trivial component $\CAL{F}_{V R}$ and one has
\be
\CAL{F}_{V R}=\CAL{A}_{R,V}-\CAL{A}_{V,R}\, .
\ee
We define the electric invariant as follows
\be
\CAL{E}=-\frac{1}{2}e^{\mu\nu}\CAL{F}_{\mu\nu}=-\CAL{F}_{V R}\, ,
\ee
where $e^{\mu\nu}$ is a $2D$  unit antisymmetric tensor.
The sign of $\mathcal E$ is defined using the orientation of the two-dimensional orbit space. The action is understood on the corresponding electric branch; the quadratic invariant alone determines only $\CAL{E}^2$.

Integration over angle variables in both integrals in the action \eqref{NLM}  gives the same angular volume  factor $\Omega_{D-1}$. Denote the reduced version of the action by $\CAL{W}=W/\Omega_{D-1}$, then one has
\be
\CAL{W}=\int d^{2}{X} R^{D-1} \Big[L(\CAL{E})- \CAL{J}^{\mu} \CAL{A}_{\mu}\Big] \, .
\ee
As earlier we use Greek indices for the components of vectors and tensors in $2D$ sector $(V,R)$.
The field equations obtained from this reduced action by its variation over a vector potential are
\be \n{NELEQ}
\begin{split}
&\CAL{H}_{,R}
=-R^{D-1}\CAL{J}^{V}\, ,\\
& \CAL{H}_{,V}
= R^{D-1}\CAL{J}^{R}\, ,
\end{split}
\ee
where
\be
\CAL{H}=R^{D-1}h(\CAL{E})\hh h(\CAL{E})=\frac{dL}{d\CAL{E}}\, .
\ee
The compatibility condition for the equations \eqref{NELEQ} is obtained by
taking a $V$ derivative of the first equation and an $R$ derivative of
the second one. This gives
\be
\partial_V\!\left(R^{D-1}\CAL{J}^{V}\right)
+
\partial_R\!\left(R^{D-1}\CAL{J}^{R}\right)= 0.
\n{CONS}
\ee
Thus, the dimensionally reduced current $j^{\mu}=R^{D-1}\CAL{J}^{\mu}$ is conserved $j^{\mu}_{;\mu}=0$.

To establish the connection between these equations and the QTG problem, let us make the following observation.
Consider a spherically symmetric scalar function $B(V,R)$ defined in the flat $(D+1)$-dimensional spacetime $M^{D+1}$. Writing $V=T+R$ and restricting $B(T+R,R)$ to a hypersurface of constant $T$, one obtains a function of the single variable $R$.
On the equatorial plane $\Pi$, defined by $X_D=0$, one has $R|_{\Pi}=r$ and hence
\be
B(T+R,R)\big|_{\Pi}=B(T+r,r)=B(v,r),
\ee
where $v=T+r$. Thus, a spherically symmetric scalar field in $M^{D+1}$ naturally induces, through its restriction to $\Pi$, a spherically symmetric scalar field on $M^D$.

Making use of this correspondence, we define the current appearing on
the right-hand side of \eqref{NELEQ} by
\be\n{CURR}
\begin{split}
\cal{J}^V &= \frac{2\kappa}{D-2}\,
      \frac{\tau(V,R)}{R^{D-1}}\,,\\
\CAL{J}^R &= \frac{2\kappa}{D-2}\,
      \frac{\sigma(V,R)}{R^{D-1}}\,,
\end{split}
\ee
where $\tau(V,R)$ and $\sigma(V,R)$ are smooth extensions of the
functions $\tau(v,r)$ and $\sigma(v,r)$ from the hypersurface
$\Pi$ to the ambient flat $(D+1)$-dimensional spacetime $M^{D+1}$.

With this definition, the current conservation law
\eqref{CONS} takes the form
\be \n{CONS1}
\tau_{,V}+\sigma_{,R}=0\,.
\ee
Equations \eqref{NELEQ} being restricted to $\Pi$ take the form
\be \n{HHVR1}
H_{,v}=\frac{2\kappa}{D-2}\sigma
\hh
H_{,r}=-\frac{2\kappa}{D-2}\tau\,,
\ee
where $H(v,r)=\mathcal{H}(V,R)\big|_{\Pi}$, and reproduce equations \eqref{HHVR0} of QTG.
It is easy to see that \eqref{CONS1} on $\Pi$ reduces to the first equation in \eqref{EEQQ}.

Let us summarize the construction. To solve the QTG field equations coupled
to the stress-energy tensor \eqref{TTTT}--\eqref{SETT}, one first
considers a spherically symmetric nonlinear (auxiliary) electric field
$\mathcal{E}$ in the flat spacetime $M^{D+1}$ with a Lagrangian density
$L(\mathcal{E})$ satisfying
\[
\frac{dL(\mathcal{E})}{d\mathcal{E}}=h(\mathcal{E}),
\]
where the function $h$ specifies the particular QTG model.

After solving Eqs.~\eqref{NELEQ} with the current \eqref{CURR}, the
resulting function $\mathcal{H}$ is restricted to the hyperplane $\Pi$,
and the identifications
\be
\mathcal{E}\big|_{\Pi}=p,\qquad
\big(\mathcal{H}/R^{D-1}\big)\big|_{\Pi}=h(p)
\ee
are made.

The metric function $f(v,r)$ in the metric \eqref{NFR} is then obtained
by inverting the relation $h=h(p)$ to determine $p=p(h)$ and
substituting
\be
f=1-r^2p(v,r).
\ee
Finally, choosing the gauge $N=1$ yields the desired solution of the QTG
field equations with the matter source
\eqref{TTTT}--\eqref{SETT}.

\section{Examples}\label{Sec4}

It should be emphasized that, up to this point, the matter source has
been treated as a prescribed external distribution constrained only by
the assumed ansatz. In fact, stress-energy tensors of the form
\eqref{TTTT}--\eqref{SETT} arise naturally in a broad class of field
theories, including Maxwell electrodynamics, nonlinear
electrodynamics, and Yang--Mills theory. In  this section we show
this explicitly.

\subsection{Non-linear electrodynamics}

In this subsection, we demonstrate that the stress--energy tensor of a Maxwell field, as well as of its nonlinear generalization generated by a spherically symmetric current distribution, is of the form \eqref{TTTT}--\eqref{SETT}.

Let us consider a curved $D$-dimensional spacetime $\CAL{M}^D$ and an electromagnetic field $A_A$ defined on it. We emphasize that this field is distinct from the auxiliary gauge field introduced in the previous section. We assume that the electromagnetic field is coupled to QTG and that its stress-energy tensor acts as a source for the gravitational field.
We choose the action  for this field in the form
\footnote{
For useful references on this subject see e.g.
\cite{BornInfeld, Ketov:2001dq, Kerner:2001qq, Sorokin:2021tge, Yang:2023BI, SingularitiesNED}.
}
\begin{equation}
 S_\ins{m}=-\frac{1}{16\pi}\int d^D X\sqrt{-g}\,{L}(\CAL{F})-\int d^D X\sqrt{-g}\,A_A  J^A \, .
 \label{NED_action}
\end{equation}
Here   ${F}_{AB}=\pa_A {A}_B-\pa_B {A}_A$,
$\CAL{F}=F_{AB}F^{AB}$ and  $J^A$  is the current.
The Maxwell theory is recovered for
${L}(\CAL{F})=\CAL{F}$.
Variation with respect to $A_A$ gives
\begin{equation}\label{NED_equation1}
\nabla_B\left(K F^{BA}\right)=4\pi J^A
\hh K(\CAL{F})=\frac{d {L}}{d \CAL{F}} \, .
\ee
The stress-energy tensor is
\begin{equation}\n{EMSET}
 T_{AB}^\ins{NED}=\frac{1}{4\pi}\left(
 K F_{AC}F_B{}^C-\frac{1}{4}g_{AB}{L}
 \right) .
\end{equation}

For a spherically symmetric electric field in the spacetime $\CAL{M}^D$, written in the $(v,r)$ coordinates of Eq.~\eqref{NFR}, the vectors $A^A$ and $J^A$ have nonvanishing components only in the $v$ and $r$ directions, while the field-strength tensor has only one independent nonvanishing component
\be
E=F_{vr}=-F_{rv}\, .
\ee
The electromagnetic invariant becomes
$\CAL{F}=-{2E^2}/{N^2}$. The stress-energy tensor takes the form \eqref{TTTT}-\eqref{SETT} with
\be
\tau=\frac{r^{D-2}}{4\pi}
\left(
-\frac{K E^2}{N^2}-\frac14 L
\right)
\hh
\mathcal{T}=-\frac{(D-2)r^{D-2}}{16\pi} L.
\ee
Since $\mathcal{T}_{rr}=0$, the QTG equations imply $N=1$. If the electric charge is not constant but depends on $v$, the stress-energy tensor \eqref{EMSET} is no longer conserved. To ensure its conservation, one must supplement it with a term proportional to $\sigma v_{,\mu}v_{,\nu}$, which accounts for the contribution of the charged matter flux associated with the electric current.

\subsection{Yang-Mills fields}

For the Yang--Mills  in a $D-$dimensional spacetime \eqref{GO} interacting with an external color current, we use the action
\begin{equation}\n{YMA}
\begin{split}
 S_{YM}
 &=
 -\frac{1}{4g_{YM}^{2}}
 \int d^{D}x\,\sqrt{-g}\,
 \kappa_{mn}F^{m}_{BC}F^{n\,BC}\\
& +
 \int d^Dx\,\sqrt{-g}\,\kappa_{mn}A^m_BJ^{n B}.
 \end{split}
\end{equation}
Here $g_{\rm YM}$ denotes the Yang--Mills coupling constant, and $T_m$ are the generators of the Lie algebra $\mathfrak{g}$ associated with the Yang--Mills gauge group $G$
\begin{equation}
 [T_m,T_n]=C^{k}{}_{mn}T_k,
\end{equation}
and write
\begin{equation}
 A_B=A_B^mT_m,
 \qquad
 F_{BC}=F^m{}_{BC}T_m.
\end{equation}
The Yang--Mills field strength in the matrix form is
\be
 F_{BC}=
 \partial_B A_C-\partial_CA_B+[A_B,A_C],
 \label{Fmatrixdef}\
\ee
In matrix notation,
\begin{equation}
 D_BX=\nabla_BX+[A_B,X].
 \label{DAmatrixdef}
\end{equation}

Variation of action \eqref{YMA}  with respect to the gauge potential gives
\begin{equation}
 D_B F^{m\,BC}=-g_{YM}^{2}J^{mC}.
 \label{YMequationcurrent}
\end{equation}
Gauge covariance of this equation requires the covariant conservation
law
\begin{equation}
 D_CJ^{mC}=0.
 \label{currentconservation}
\end{equation}
The Yang--Mills-field contribution to the stress--energy tensor is
\be
T_{BC}=\frac1{g_{\rm YM}^2}\kappa_{mn}
\left( F^m_{BD}F^{n}{}_C{}^{D} -\frac14 g_{BC}F^m_{DE}F^{n\,DE} \right)\, .
\ee
In Yang–Mills theory with group $G$, $\kappa_{mn}$ is an invariant metric on the Lie algebra $\mathfrak g$ (or an invariant bilinear form). When it is specifically constructed from the adjoint representation, it is called the Killing form.

In the presence of the external current, the Yang--Mills stress
tensor is not conserved separately. Using \eqref{YMequationcurrent}, one
finds
\begin{equation}
 \nabla_B T^{(YM) B}{}_{C}
 =-\kappa_{mn}F^m{}_{CB}J^{nB}\, .
 \label{fullconservation}
\end{equation}
It is worth emphasizing that the dependence of the stress-energy tensor on the gauge group enters only through the invariant metric $\kappa_{mn}$ and the normalization of the generators. We do not specify the Yang--Mills gauge group $G$, assuming only that it admits nontrivial spherically symmetric gauge field configurations. The general classification of gauge groups supporting such configurations, together with the corresponding existence theorems based on invariant connections over homogeneous spaces, can be found, for example, in Refs.~ \cite{ForgacsManton1980,GuHu1981,Brodbeck1996,HarnadShniderVinet1980,OliynykKunzle2002a,OliynykKunzle2002b}.

An important class of spherically symmetric Yang--Mills configurations is characterized by the vanishing of the mixed components of the field strength,
$F_{\mu i}=0$.
This class includes both purely electric configurations, for which
$F_{ij}=0$,
and purely magnetic configurations, for which
$F_{\mu\nu}=0$ .

In the absence of a color current and when the mixed components vanish, $F_{\mu i}=0$, the stress--energy tensor assumes the form
\be\n{YMSET}
\begin{split}
&T_{\mu\nu} = \frac1{g_{\rm YM}^2} \gamma_{\mu\nu}
\left(\frac14X-\frac{Z}{4r^4}\right)\hh
T_{\mu i}=0\, ,\\
&T^i{}_j=\frac{\delta^i{}_j}{g_{\rm YM}^2}
\left[
-\frac14X
+\left(\frac1{D-2}-\frac14\right)\frac{Z}{r^4}
\right].
\end{split}
\ee
Here we introduce the invariants
\be
\begin{split}
&X=\kappa_{mn}F^m_{\mu\nu}F^{n\,\mu\nu}\, ,\\
& Z=\omega^{ik}\omega^{jl} \kappa_{mn} F^m_{ij}F^n_{kl}.
\end{split}
\ee
Let us note that  the \emph{functional form} of the stress tensor is universal.
The reduced components of the stress-energy tensor  are
\[
\tau=\frac{r^{D-2}}{4g_{\rm YM}^2}\left(X-\frac Z{r^4}\right),
\]
\[
\CAL{T}=\frac{r^{D-2}}{4g_{\rm YM}^2}
\left[-(D-2)X+(6-D)\frac Z{r^4}\right],\qquad \sigma=0.
\]

It is straightforward to verify that the Yang--Mills stress--energy tensor \eqref{YMSET} is of the form \eqref{TTTT}--\eqref{SETT} with $\sigma=0$.
When a null Yang--Mills current is present, the total stress--energy tensor acquires an additional null-fluid contribution proportional to $v_{,B}v_{,C}$, corresponding to a nonvanishing function $\sigma$.

The form of the Yang--Mills stress--energy tensor \eqref{YMSET} implies that, in the absence of Yang--Mills currents, the corresponding solution of the QTG field equations satisfies a generalized Birkhoff theorem and therefore admits a Killing vector~\cite{Frolov:2026tft}. Moreover, in the null coordinates $(v,r)$ introduced in \eqref{NFR}, one may choose the gauge $N=1$.

\section{Discussion}\label{Sec5}

We have developed a modified double-copy formulation of
quasitopological gravity in the presence of matter sources. The
construction replaces the direct solution of the nonlinear gravitational
field equations in a curved spacetime $\CAL{M}^D$ by the solution of gauge-field equations in an
auxiliary flat $(D+1)$-dimensional spacetime. The gravitational metric
is then reconstructed from the gauge-field solution through the
generating function defining the QTG theory.

For spherically symmetric configurations, the field equations reduce to
a remarkably simple system. The dependence on the underlying gravity
theory is encoded entirely in the generating function $h(p)$, while the matter
sector enters only through its stress--energy tensor. We impose a special ansatz on its form.
We demonstrated
that Maxwell theory, nonlinear electrodynamics, and Yang--Mills fields
all fit naturally into this framework. In particular, for Yang--Mills
fields the only dependence on the gauge group is through the invariant
Lie-algebra metric and the normalization of the generators, provided
the gauge group admits nontrivial spherically symmetric
configurations.

An important result is that matter sources whose stress--energy tensor
is of the form considered in this paper and does not contain null-matter  fluxes
 satisfy the conditions required
for the generalized Birkhoff theorem. 
Consequently, in the absence of
null currents the corresponding solutions possess a Killing vector. When null charged matter is present, the
formalism naturally yields Vaidya-type solutions describing the
time-dependent evolution of the mass and charge.

The present approach opens several directions for future work. These
include rotating solutions, non-Abelian regular black holes, dynamical
collapse, black-hole evaporation, and cosmological models in
quasitopological gravity. It would also be interesting to investigate
whether the modified double-copy construction can be extended beyond
spherical symmetry and applied to more general classes of higher-curvature theories.

\acknowledgments

This work was supported by the Natural Sciences and Engineering Research Council of Canada (NSERC). The author also gratefully acknowledges financial support from the Killam Trust.
The author is grateful to Chulmoon Yoo (Nagoya University) and Andrei Zelnikov (University of Alberta) for valuable discussions.


%



\end{document}